\documentclass [12pt]{article}
\begin{document}
\title{ Replies to Comments on "Tetraquarks  as Diquark-Antidiquark Bound Systems" }
\author
{M. Monemzadeh\thanks{monem@kashanu.ac.ir}\\
N. Tazimi \thanks{nt$_{-}$physics@yahoo.com}\\
P. Sadeghi \thanks{ph.parva@yahoo.com}\\
\it\small{{Department of Physics, University of  Kashan, Kashan, Iran.}}}
\date{}

\maketitle
\begin{abstract}
The present paper is written in response to the comments of M.R. Hadizadeh on our original paper "Tetraquarks as diquark-antidiquark bound systems" [Phys. Lett. B \textbf{741}, 124 (2015)]. We present our clarifications on his arguments against the accuracy of the procedure and results of the study. The arguments turn out to stem mainly from the author's misunderstanding of the issues discussed in the original paper.
 \end{abstract}

\section{Introduction}
In a recent paper titled "Tetraquarks as Diquark-Antidiquark Systems" we studied the tetraquark systems containing c quark. After revisions and replies to the reviewers' miscellaneous comments, the manuscript was ultimately published in PLB \cite{1}. However, as research is generative, every discovery brings about further questions. Publication of  "Tetraquarks as Diquark-Antidiquark Systems" was followed by \textit{Comments on "Tetraquarks as Diquark-Antidiquark Systems"} by M.R. Hadizadeh \cite{2}. In the first place, we would like to express our gratitude for his meticulous scrutiny into the article. In the second place, we are totally dedicated to presenting our replies to the comments in order to compromise on any technical objections as well as misunderstandings apparent within the comments:\\

	1) As for the coefficient $\sqrt{\frac{\pi}{2}}  $ in equations 1, 4, 9 and 11, it must be pointed out that it is merely a matter of human mistake while typing. The correct coefficient  is ${\frac{1}{4\pi}}  $ . It is emphasized that ${\frac{1}{4\pi}}  $ was rightly used in the Fortran code in our calculations, so the tetraquark masses were derived based on inserting ${\frac{1}{4\pi}}  $  in the calculations. Indeed, if we had used $\sqrt{\frac{\pi}{2}}  $ rather than ${\frac{1}{4\pi}}  $  , we would have arrived at different and unacceptable mass measures.\\

	 2)  The coupling constant we used in Ebert potential has been calculated via (2) in Ref. \cite{2}.  However, since we made reference to Ebert et al. \cite{3}, we didn't include the equation in the body of the article. In our recent work \cite{4}, also we employed this coupling constant in meson systems.\\

	3)  The following  linear variable change is used to transfer the  integration interval of $r \prime$ from $[0, r_{max}]$ to $[-1,+1]$:
\begin{equation}
r =r_{max} \frac{1+x}{2} 
\label{5}
\end{equation}
 As the author Hadizadeh points out, he has carried out the calculations in configuration space using linear and hyperbolic variable changes, but we conducted our calculations using linear variable changes.\\
In configuration space, r-cutoff does not significantly influence the outcome. The confining term in the potential brings about divergence and thus, we supposed V as constant and equal to zero for any r exceeding r-cutoff. In two-body calculations, r-cutoff will have no significant effects in configuration space. (In Table 2 of [2], the author Hadizadeh presents the fact that varying r-cutoff leads to only minor changes in mass.)
However, for two- and three-body calculations in momentum space, this renormalization method is not applicable, but it works very well with two-body calculations \emph{in configuration space} in long ranges. In our previous works, including [5] (in which M.R. Hadizadeh was an author), we also successfully employed this method to solve the integral Lippman-Schwinger equation and verify the validity of the code and numerical calculations, and we calculated the binding energy of  $t\bar{t}  $,$b\bar{b}  $, and $c\bar{c}  $ systems with high accuracy \cite{5}.
Furthermore, besides r-cutoff another parameter called $r^{'}$-cutoff is involved in the calculations of two-body systems in configuration space. This parameter can also influence the results.\\

4) As the author points out, this formalism is applicable to $ L=0 $  states. Some of the details of our methodology in \cite{1} were not given elaborately and this may have triggered misunderstanding. As the author holds, we had used a spin-independent unrelativistic potential and our results definitely apply to the ground state. Evidently, more accurate results are attainable by considering spin effects in the potential.

 However, we found two energies for $S \bar{A}$ state and three for $ A \bar{A}$. These were well consistent with $J \neq 0$ state. Hence, we conjecture and suggest that these findings can potentially correspond to $J \neq 0$ state and are in agreement with Ebert’s \cite{3} results. \\

For  $A \bar{A}$ state, there are three  angular momentums, i.e.   $ J=0^{++},1^{+-},2^{++} $  and $J=1^{+-},1^{+-}$ for $ \frac{S \bar{A}\pm A \bar{S}}{\sqrt{2}} $ states. Thus, utilizing\\
$$ J_{tot}=J_{D}+J_{\bar{D}}=(L+S)_{D}+(L+S)_{\bar{D}}    $$
$$ if ({l=0})  \Longrightarrow J_{tot}= S_{D}+S_{\bar{D}}$$\\
and the parameters $M, \xi ,\zeta $ in Table 1 of [1], we have calculated different states of $ J $  by considering the system's reduced mass. For example, we obtained two energies $ E_{1} $ and $ E_{2} $ ($E_{1}>E_{2}$ ) yielding  $ \lambda=1 $  in  $ \frac{S \bar{A}\pm A \bar{S}}{\sqrt{2}} $ for this system's  reduced mass. As a result,  $ E_{1} $ is supposed to correspond to $ \frac{S \bar{A}+ A \bar{S}}{\sqrt{2}} $ and $ E_{2} $ to $ \frac{S \bar{A}- A \bar{S}}{\sqrt{2}} $. \\

For $A\bar{A}$ state, we obtained 3 energies $ E_{1} $,$ E_{2} $  and $ E_{3} $  ($E_{1}>E_{2}>E_{3}$) yielding $ \lambda=1 $. Thus, we suggest that the highest energy $ E_{1} $ corresponds to $ J=2 $, the lowest energy $ E_{3} $ to $ J=0$, and $ E_{2} $ to $ J=1$.\\

However, by solving  Lippmann-Schwinger equation, we obtained tetraquark masses in ground state and  propose tetraquark masses for non-zero total angular momentum. As stated in the conclusion of \cite{1}, the method enjoys being capable of reducing a complicated four-body problem to a simpler two-body problem. Solving the four-body problem via the relativistic method for tetraquarks containing light quarks will entail far more complicated calculations.\\

Taking all our replies to the disputes against [1] into account, we reject the author's statement that all of the results published in recent letter by M. Monemzadeh et al. [1], for masses of tetraquarks with total angular momentum equal to zero, are incorrect and the letter has serious numerical problems. Instead, we assert that our results agree with other already published results and our method, being simpler, less time-consuming and hence more economical, can be a good replacement for complicated relativistic methods.

\end{document}